# A Photonic-CXL Memory Appliance for Scalable KV Cache Management in LLM Inference

Jing Ding
Machine Learning
Marvell Technology
Santa Clara, USA
jingd@marvell.com

Yash Nishant
Machine Learning
Marvell Technology
Santa Clara, USA
ynishant@marvell.com

Chandrish Ambati
Machine Learning
Marvell Technology
Santa Clara, USA
cambati@marvell.com

Jyothsna Kamati
Design Verification
Marvell Technology
Santa Clara, USA
jkamati@marvell.com

Trung Diep
Machine Learning
Marvell Technology
Santa Clara, USA
trungd@marvell.com

***Abstract*— LLM inference at scale faces a memory wall: the KV cache demands tens of terabytes at hundreds of gigabytes per second, yet no current memory tier delivers both at once. Characterization across multi-generation GPU systems with various LLaMA models shows host memory retrieval achieves up to 100x speedup over re-computation but supports only tens of concurrent long-context users. Electrical CXL pooling theoretically bridges this gap, but switch latency, cable reach limits, and power-scaling issues prevent practical TB-scale deployments. We present the Marvell® Photonic Fabric™ (PF™) Memory Appliance, a photonic-CXL hybrid architecture replacing electrical switches with a passive fiber shuffle to deliver 32 TB shared memory across 16 hosts via a switch-free full-crossbar topology. Emulation results demonstrate over 50 percent latency reduction versus electrical CXL pools. Simulation results show that the PF Memory Appliance eliminates cache eviction cliffs by improving time-to-first-token by 6.6x for multi-turn conversations workloads.**

## I. Introduction

The rapid deployment of Large Language Models (LLMs) in production systems has created unprecedented challenges in memory management, particularly for Key-Value (KV) cache placement. As models scale to support context windows exceeding 1 million tokens—exemplified by Claude Opus, Gemini 3, GPT-5.4, LLaMA 4 Maverick [1, 2, 3, 4]—the memory requirements for KV cache have become a critical bottleneck in achieving cost-effective and performant inference at scale.

Consider a LLaMA 70B parameter model serving enterprise applications. For each token in the context, the KV cache requires approximately 328KB (assuming BF16 precision with 80 layers, 8 KV heads, and 128-dimensional head size). In typical enterprise deployments, this translates to substantial memory requirements. Customer Support Systems typically begin each conversation with a 2,000-token system prompt containing company policies and documentation, followed by user-specific context. After 10–15 exchanges, the total context easily exceeds 10,000 tokens, with storage of ~3.3 GB per user. With 3.5 billion daily active users at Meta and 1.3 million GPUs in datacenter infrastructures, each GPU would require 8.9 TB for serving all users.

Enterprise applications at scale face even more severe memory challenges due to their inherently long contexts and high concurrency requirements. Legal document analysis systems must process extensive contracts alongside vast regulatory frameworks and case precedents, with each session requiring tens of gigabytes of KV cache memory. Medical diagnosis platforms combine complete patient histories with comprehensive medical knowledge bases and clinical guidelines, quickly exhausting available GPU memory when serving multiple healthcare providers simultaneously. Educational platforms delivering personalized tutoring must maintain extensive context including curriculum materials, learning history, and performance analytics for thousands of concurrent students, pushing memory requirements into the multi-terabyte range. These domain-specific applications demonstrate that KV cache management is a fundamental barrier to deploying LLMs in enterprise settings where long context and high concurrency are non-negotiable requirements.

Current inference frameworks employ hierarchical memory management to address these challenges. NVIDIA's Dynamo [5], vLLM [6] with LMCache [7] integration, and SGLang (HiCache) [8] implement multi-tier caching strategies spanning GPU HBM, host DRAM, local SSDs, and network-attached storage. While host memory offloading effectively preserves GPU compute resources and accelerates inference for long-context workloads, it faces fundamental capacity constraints. Modern servers equipped with up to 2 TB of DRAM can support only a limited number of active users before exhausting available memory. SSD-based caching offers virtually unlimited capacity but introduces a critical bandwidth bottleneck—10 to 15× lower than CPU-to-GPU bandwidth. Consequently, loading KV cache from SSDs can be even slower than recomputing it entirely, making it viable mostly for offline processing scenarios where latency sensitivity is not critical.

Meanwhile, disaggregated inference architectures—which separate the compute-bound prefill phase from the memory-bandwidth-bound decode phase onto distinct GPU workers [9, 10, 11]—further exacerbate the data movement challenge. KV tensors produced during prefill must be delivered to decode workers before token generation can begin, and this transfer latency increasingly dominates time-to-first-token (TTFT) at long context lengths. Existing transfer mechanisms—from CPU-driven RDMA to GPUDirect RDMA [12] to tiered KV offloading via NIXL [13]—each leave components on the critical path that limit throughput and introduce latency variability. CXL shared memory has emerged as a promising alternative, enabling byte-addressable access across multiple hosts without involving the network stack [14]. Recent work including TraCT [15] and Beluga [16] demonstrates CXL as both a KV transfer substrate and a rack-wide prefix cache. However, electrical CXL deployments face fundamental scaling constraints: switch-induced latency of 70–500 ns per hop, multi-meter cable reach limits requiring power-hungry retimers, and super-linear power scaling that prevents practical TB-scale memory pooling.

In this paper, we present a comprehensive characterization of KV cache retrieval efficiency across the memory hierarchy, evaluating offloading across A100, H100, and H200 systems with models from 8B to 405B parameters and contexts up to 4M tokens using repeated request. Our findings reveal that retrieving KV cache from host memory achieves up to 100× speedup over GPU re-computation but is severely capacity-constrained, while retrieving KV cache from SSD delivers limited speedups (maximum 9.7×) due to bandwidth bottlenecks. These results quantify the critical requirement: tens of terabytes of capacity at over 100 GB/s bandwidth—capabilities that current systems cannot simultaneously deliver.

To address this gap, we present the Marvell® Photonic Fabric™ (PF™) Memory Appliance, a novel photonic fabric-CXL hybrid architecture. By replacing electrical switches with a passive optical fiber shuffle [17], the PF Memory Appliance delivers 32 TB capacity with 128 GB/s unidirectional bandwidth per host, enabling up to 16 hosts to share pooled memory without interconnect bottlenecks. We evaluate the PF Memory Appliance design through emulation on the Siemens Veloce Strato-M platform, demonstrating over 50% latency reduction compared to electrical CXL switched memory pools [16]. End-to-end serving simulations confirm that PF Memory Appliance eliminates KV cache eviction cliffs, improving TTFT by 6.6× for multi-turn conversations workload.

Our contributions are as follows:

An empirical characterization of the memory wall in LLM inference, quantifying the capacity-bandwidth gap across GPU HBM, host DRAM, and SSD tiers.

The PF Memory Appliance architecture, a photonic-CXL hybrid achieving 32 TB shared memory through a switch-free optical fiber shuffle, with bandwidth and latency evaluated through hardware emulation.

End-to-end serving evaluation demonstrating that PF Memory Appliance-augmented systems sustain flat TTFT as workload scales to 300 concurrent conversations, achieving 6.6× improvement over eviction-degraded baselines.

The remainder of this paper is organized as follows. Section II provides background on our empirical characterization of the memory wall and disaggregated inference transfer mechanisms. Section III introduces the PF Memory Appliance architecture. Section IV presents the emulation methodology and hardware evaluation results. Section V describes software integration with production inference frameworks. Section VI evaluates end-to-end serving performance with simulations. Section VII discusses related work. Section VIII addresses limitations and future work. Section IX concludes.

## II. Background and Motivation

### A. Empirical Characterization of the Memory Wall

To quantify the capacity-bandwidth gap and establish design requirements for next-generation memory architectures, we conduct a systematic characterization of KV cache retrieval performance across the memory hierarchy. Our evaluation spans A100, H100, and H200 GPU systems with models from 8B to 405B parameters and contexts up to 4M tokens.

#### 1) Repeated Request Benchmark

The benchmark isolates KV cache retrieval performance by measuring load latency under ideal conditions (100% cache hit rate) with identical repeated requests. The experimental workflow consists of three phases:

Phase 1: Input creation and initial cache population. We generate test inputs with configurable batch sizes (1–32) and prompt lengths (1K–4M tokens), submit the initial request to the model for prefill computation, during which the model generates and stores the complete KV cache.

Phase 2: Cache offloading. For the host memory path, cache is offloaded to system DRAM for fast retrieval with limited capacity. For the disk storage path, cache persisted to SSD for large capacity with slower retrieval.

Phase 3: Retrieval performance measurement. We measure memory retrieval time ($T_{memory}$) to load KV cache from host DRAM, disk retrieval time ($T_{disk}$) to load KV cache from SSD, and a no-cache baseline ($T_{compute}$) for full KV re-computation. The speedup of retrieving KV cache from host memory is calculated as $T_{compute} / T_{memory}$, while the speedup from SSD is $T_{compute} / T_{disk}$.

The benchmark evaluates five model variants: LLaMA-8B (single GPU), LLaMA-70B (2×H200, TP=2), LLaMA-405B (8×H200, TP=8, BF16), and extended-context MoE variants LLaMA Maverick and LLaMA Scout (8×H200, TP=8, BF16).

#### 2) Experimental Setup

We run vLLM with LMCache enabled to offload and retrieve KV blocks across requests. We compare tier configurations that offload KV blocks to host DRAM and local storage when HBM is insufficient. We report time-to-first-token (TTFT) as the primary user-facing latency metric.

Our evaluation uses a local A100 server and cloud instances on Runpod with H100/H200 GPUs. The A100 testbed includes one NVIDIA A100 (80 GB HBM2e) connected to a dual-socket

Intel Xeon server over PCIe 4.0 ×16 (~32 GB/s unidirectional) with 512 GB system DRAM and 7 TB local disk. The H100/H200 testbeds provide PCIe 5.0 ×16 (~64 GB/s unidirectional), with multi-GPU nodes (up to eight H100 or H200) providing up to 2 TB host DRAM.

*3) Results*

TABLE I summarizes the KV cache retrieval speedup ranges across all model-platform configurations. All configurations use batch sizes {1, 2, 8, 16, 32}. Standard context lengths span 1K–100K tokens; Maverick extends to 1M; Scout to 4M. The experimental results demonstrate that host memory-based KV cache retrieval consistently provides substantial performance improvements over baseline re-computation, while disk-based caching shows more nuanced benefits depending on workload, model size, and deployed systems.

TABLE I. TESTING MATRIX AND PERFORMANCE SPEEDUP RANGES FOR REPEATED REQUEST BENCHMARK. ALL CONFIGURATIONS USE BATCH SIZES {1, 2, 8, 16, 32}. STANDARD CONTEXT LENGTHS: 1K--100K TOKENS; MAVERICK EXTENDS TO 1M; SCOUT TO 4M. "—" INDICATES NOT TESTED.

| Model | Platform | Host Memory | Disk Storage |
|---|---|---|---|
| LLaMA-8B | 1×A100 | 3.3×-27.5× | 0.53×-1.72× |
| LLaMA-8B | 1×H100 | 1.9×-11.8× | — |
| LLaMA-8B | 1×H200 | 1.5×-17.2× | 0.38×-1.27× |
| LLaMA-70B | 2×H200 | 2.7×-51.4× | — |
| LLaMA-405B | 8×H200 | 2.7×-100.0× | 2.1×-9.7× |
| Maverick | 8×H200 | 2.7×-23.1× | — |
| Scout | 8×H200 | 2.6×-54.6× | — |

**Host memory performance.** The magnitude of speedup correlates strongly with model size and context length. The 405B model on the 8×H200 system achieves up to 100× speedup, while the 8B model achieves 17.2× on a single H200. This exceptional performance for 405B stems from the model's substantial KV cache requirements (48 GB for 100K tokens compared to 12 GB for 8B) and the system's ability to aggregate bandwidth across eight GPUs, reaching 165 GB/s effective bandwidth.

The extended-context MoE variants (Maverick and Scout) present an unexpected pattern. Despite supporting dramatically longer contexts (1M and 4M tokens respectively), their maximum speedups (23.1× and 54.6×) remain below the 405B dense model's peak. This arises from the MoE architecture's computational efficiency: sparse expert activation reduces baseline compute requirements, thereby diminishing the relative advantage of cache retrieval.

Across all models, speedup factors exhibit consistent scaling with prompt length (Fig. 1(a) for LLaMA 405B results), validating the theoretical advantage of $O(n)$ cache retrieval over $O(n^2)$ attention computation. Speedup also increases with batch size (Fig. 1(b)), as the fixed overhead of CPU-to-GPU transfer—cache lookup, memory allocation, and PCIe transfer initialization—is amortized across more sequences per batch.

The speedup $S$ for dense models can be approximated as:

$$S = T_{compute} / T_{memory} \quad (1)$$

$$T_{compute} = b(2sP + 2s^2 n_{head} h_{head} L)/MFU \quad (2)$$

$$T_{memory} = K + bs\alpha \quad (3)$$

where $b$ is batch size, $s$ is prompt length, $P$ represents the number of parameters, $n_{head}$ is the number of attention heads, $h_{head}$ is the head dimension, $L$ is the number of transformer layers, $K$ represents fixed overhead, and $\alpha$ is the per-token loading cost. For the 405B model, attention FLOPs achieve parity with feed-forward FLOPs at approximately 20,000 tokens and dominate at longer contexts, explaining the super linear scaling of speedup with prompt length.

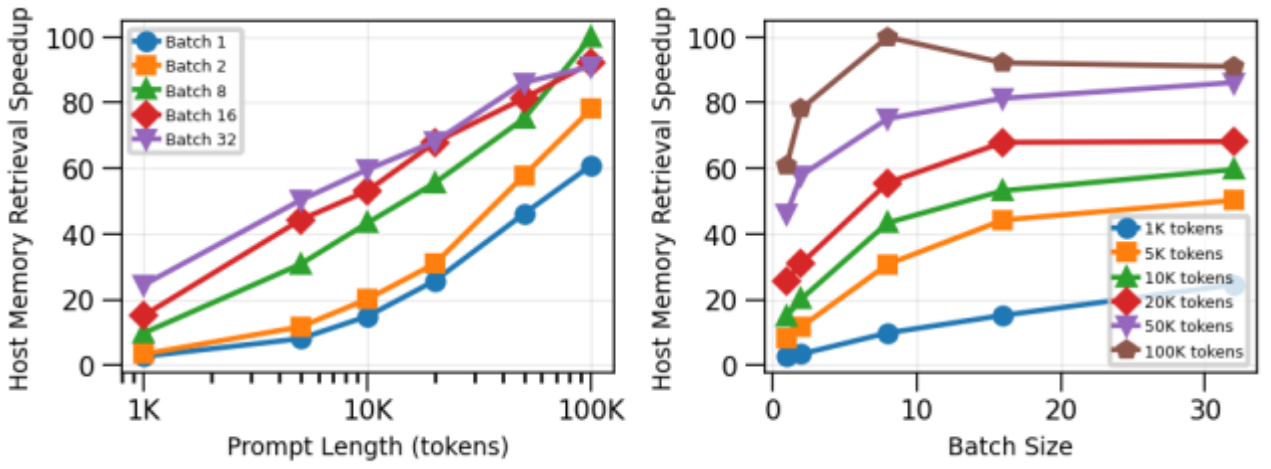

Fig. 1. KV retrieval from host memory speedup compared to GPU re-computation versus (a) Prompt Length and (b) Batch size for LLaMA **405B** model on eight H200 systems.

**Disk storage performance.** Disk-based caching diverges significantly from host memory patterns. For LLaMA-8B on the A100 system, disk retrieval provides speedups only for sequences exceeding 50K tokens, with shorter contexts actually showing performance degradation (0.38×–0.53× for sequences under 20K tokens). The LLaMA-405B model demonstrates more favorable disk caching characteristics, achieving speedups of 2.1×–9.7×, as the computation time for the 405B model exceeds disk I/O latency even with 8×H200 GPUs operating in parallel.

### *B. KV Cache transfer in Disaggregated Inference*

Disaggregated LLM serving separates the compute-bound prefill phase from the memory-bandwidth-bound decode phase onto distinct GPU workers, enabling independent scaling and hardware specialization. This separation places KV transfer on the critical path: KV tensors produced during prefill must reach the decode worker before token generation begins, and transfer latency increasingly dominates TTFT at long context lengths.

**CPU-driven RDMA.** Early disaggregated systems [9,11] transfer KV tensors over InfiniBand or RoCE with the host CPU orchestrating all data movement. The data path traverses six hops: the CPU stages KV data from GPU HBM to a pinned host DRAM bounce buffer via cudaMemcpy, then posts an RDMA verb to transmit via the NIC; the reverse occurs on the decode side. This approach incurs two intermediate host DRAM copies, four PCIe bus crossings, and places the CPU on the critical path at three serialization points, yielding 20–50 µs of added latency with high variance under network contention.

**GPUDirect RDMA.** NVIDIA GPUDirect RDMA [12] eliminates both bounce buffers by enabling the NIC to read from and write to GPU memory directly through PCIe BAR1 mappings. The CPU's role reduces to posting a single RDMA verb with GPU-address scatter-gather lists; it never issues a cudaMemcpy. This halves the PCIe crossings to two peer-to-peer transactions, reducing latency to 2–15 µs [19]. However, a well-documented read-write asymmetry limits BAR read throughput to 3.4–7 GB/s on many platforms [19], and the GPU and NIC must share a PCIe root complex. Critically, both RDMA variants are point-to-point: every prefill-to-decode handoff is a discrete network transfer, KV blocks cannot be reused across workers without retransmission, and schedulers must maintain per-node KV locality indices for cache-aware routing **[5,20]**.

**KV offload to tiered storage.** A complementary approach offloads KV blocks from GPU HBM to external storage tiers including CPU DRAM, SSD, or networked storage for later retrieval by decode workers [5, 7, 8]. Systems such as LMCache [7], Dynamo KVBM [5], and SGLang HiCache [8] manage hierarchical KV caching with asynchronous store/load and pluggable storage backends via NIXL [13]. This decouples KV production from consumption in time and enables cross-request prefix reuse, but introduces a double data movement penalty (store + load) and remains bounded by storage-tier bandwidth and network RPC latency for index coordination.

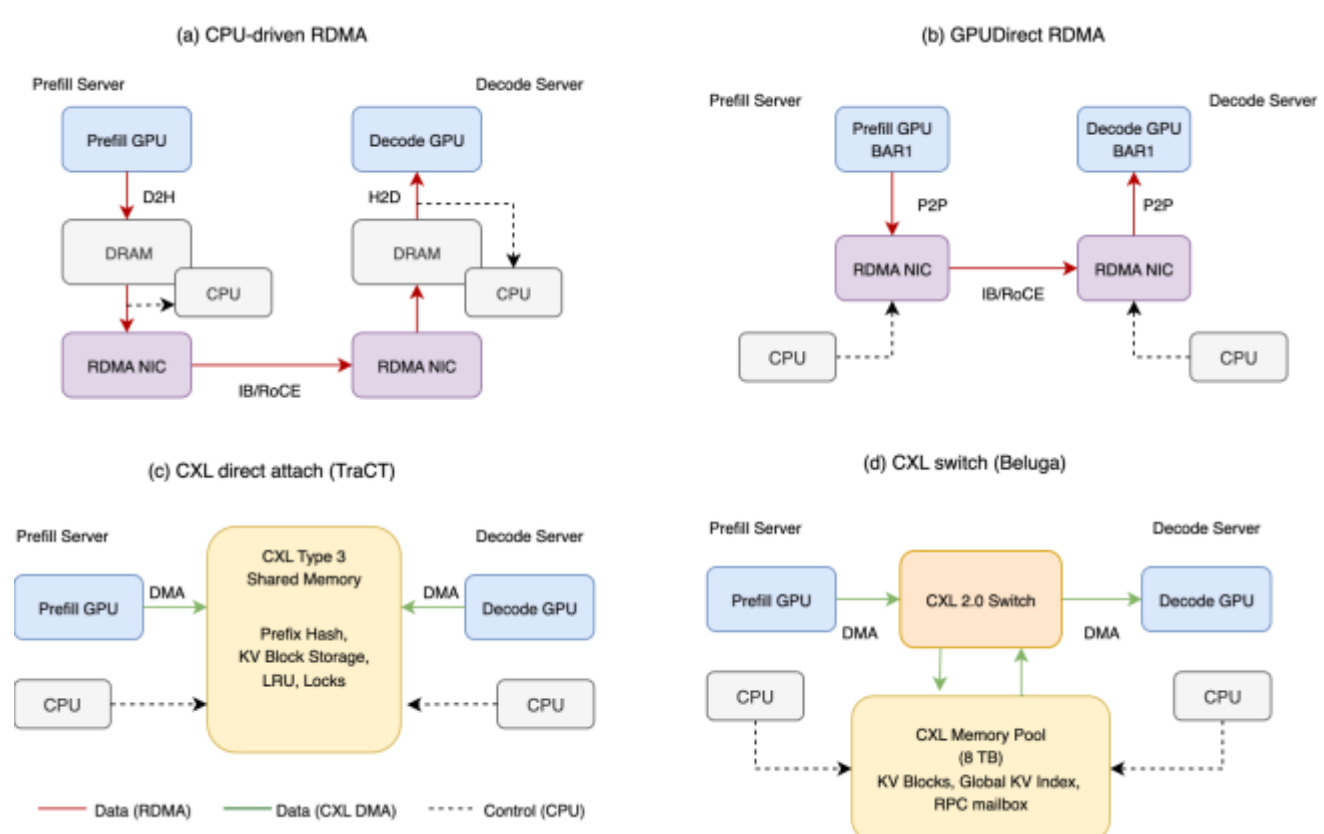

Fig. 2. Evolution of KV cache transfer data paths in disaggregated LLM inference. Solid lines represent the KV data plane; dashed lines represent CPU control operations. Each generation removes components from the critical path: (a)→(b) eliminates DRAM bounces; (b)→ (c,d) eliminates the network. CXL modes additionally enable write-once-read-many KV reuse and eliminate KV-locality-aware scheduling.

**CXL shared memory.** CXL Type-3 devices provide byte-addressable memory accessible by multiple hosts via load/store semantics, without involving the network stack [14]. Recent work demonstrates using CXL as both a KV transfer substrate and a rack-wide prefix cache. TraCT [15] maps a CXL device into all servers via DAX, registers it with CUDA for GPU DMA, and maintains a shared prefix hash table with two-tier software locks and explicit cache-line flushes for cross-node consistency on non-coherent memory. Beluga [16] scales this to 16 hosts and 8 TB capacity via a CXL 2.0 switch, replacing network RPCs with CXL-based RPCs for index coordination and enabling fine-grained access to vLLM's native block size without operation batching. Both systems eliminate the network from both data and control paths: GPU DMA writes KV blocks directly to shared CXL memory, and any decode worker reads them without retransmission. This yields write-once-read-many semantics, deterministic sub-microsecond tail latency, and eliminates KV-locality-aware scheduling entirely. On real hardware, TraCT reports up to 9.8× TTFT reduction over RDMA baselines **[15]**; Beluga achieves 89.6% TTFT reduction and 7.35× throughput improvement over RDMA-based Mooncake [16].

Fig. 2 summarizes the data path evolution. Each successive mechanism removes components from the critical path: GPUDirect eliminates DRAM bounces, CXL removes the network fabric, and CXL modes further enable write-once-read-many reuse across workers. Yet both TraCT and Beluga operate over electrical CXL links whose cable reach, switch latency, and power scaling degrade bandwidth and tail latency as pool size grows — limitations we examine next.

### C. *Bridging the Capacity-Bandwidth Gap*

Our characterization reveals a fundamental capacity and bandwidth gap in current memory hierarchies. Host memory delivers sufficient bandwidth for interactive inference but is capacity-constrained to tens of concurrent long-context users per server. SSD storage removes capacity constraints but provides inadequate bandwidth, with speedups degrading to slowdowns for contexts where I/O latency exceeds re-computation time. Production deployments serving thousands of concurrent users with long contexts—the emerging norm for enterprise and agentic applications—fall in this gap. Effective KV cache management therefore requires a memory tier combining tens-of-TB-scale capacity with bandwidth exceeding 100GB/s.

Electrical CXL can theoretically provide this combination, and the CXL-based systems reviewed above demonstrate compelling advantages over network-based KV transfer. However, their deployments expose fundamental scaling constraints inherent to electrical interconnects.

First, electrical CXL links are limited to several meters of copper reach; extending distance requires power-hungry retimers that add latency yet still confine deployments to sub-rack distances. TraCT connects hosts within a single chassis [15], and Beluga's 16-host pool relies on a co-located CXL 2.0 switch [16] — true rack-scale disaggregation remains infeasible. Second, CXL switches introduce 70–500 ns latency per hop plus variable arbitration delays under contention [14], and hierarchical topologies necessary for larger pools compound this overhead into the range where decode-phase SLO budgets are at risk. Third, switch power scales super-linearly with port count and signaling rate, creating thermal density challenges that make the hundreds of modules and multiple switch tiers needed for 32 TB pools impractical.

Scaling electrical CXL to tens of terabytes requires switch hierarchies that sacrifice the very latency and bandwidth advantages motivating CXL adoption. What is needed is an interconnect that preserves CXL's byte-addressable load/store memory semantics while eliminating electrical switching from the data path. Section III presents the PF Memory Appliance, which replaces the electrical switch fabric with a passive optical fiber shuffle to deliver 32 TB of shared CXL memory across 16 hosts without intermediate switching, distance constraints, or contention-induced latency variability.

## III. Photonic Fabric Memory Appliance Architecture

We introduce the Photonic Fabric Memory Appliance: a memory disaggregation platform that combines the capacity advantages of pooled DDR5 with the bandwidth characteristics of HBM, while leveraging photonic interconnects to eliminate the switching bottlenecks and reach limitations of electrical CXL architectures. Photonics [17] can replace the electrical switching fabric entirely, providing all-to-all connectivity without intermediate switches while maintaining CXL protocol compatibility at the host interface.

### A. System Overview

The PF Memory Appliance is a memory device built using 16 Photonic Fabric Memory Modules, providing 32 TBs of shared DDR5 memory capacity with a unified address space and 1.125 TBs of HBM3E configurable as either a write-through cache or additional pooled memory. Each PF Memory Module is a CXL 3.1 Type 3 device over a PCIe Gen6 interface, enabling general-purpose or accelerated servers to incorporate Photonic Fabric connectivity while maintaining standard software compatibility.

Each PF Memory Module is a 2.5D electro-optical multi-chip module consisting of 7.2 Tbps of Photonic Fabric-based optical connectivity, two HBM3E stacks (36GB each), eight DDR5 interfaces addressing up to 2 TB of DDR5 memory, an optical Photonic Integrated Circuit (PIC) interposer, and a Fiber Array Unit. The PF module is a memory fabric and switch ASIC packaged in a 2.5D electro-optical system-in-package (SiP) with the HBM3E stacks mounted on the PIC interposer, enabling tight integration between optical and memory subsystems.

The architectural innovation lies in the PF Memory Appliance's interconnection topology [18]. Fig. 3 presents the overall PF Memory Appliance architecture. The 16 PF Memory Modules connect to 16 front-panel optical ports through a 16×16 passive fiber shuffle providing all-to-all connectivity. This creates 256 optical paths forming a full-mesh topology, enabling any processor or server connected to any optical port to directly access any of the 16 PF Memory Modules without intermediate switching. Through this design, the PF Memory Appliance supports up to 16 hosts, with each host having direct paths to every memory module. The total optical switching bandwidth reaches 115 Tbps across the appliance, with each port delivering 7.2 Tbps of bidirectional bandwidth.

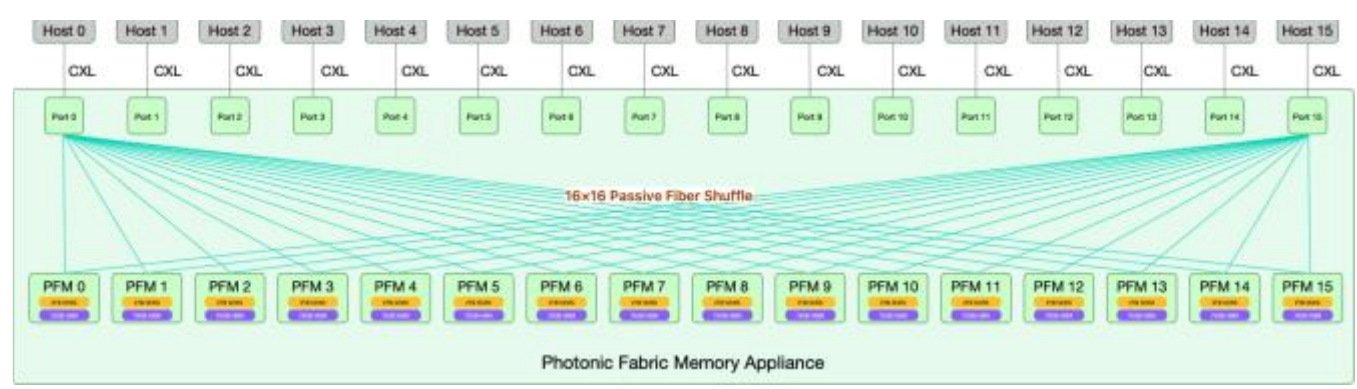

Fig. 3. PF Memory Appliance architecture: 256 optical paths creating full-mesh connectivity connecting up to 16 host servers without intermediate switching.

The passive fiber shuffle represents a fundamental departure from traditional switched architectures. Rather than routing memory transactions through active switch elements that introduce arbitration delays and potential congestion, the fiber shuffle provides dedicated optical paths between every host-memory pair. This topology eliminates switch-induced latency entirely while enabling truly concurrent access—all 16 hosts can simultaneously access different memory modules without contention in the interconnect fabric.

The PF Memory Appliance connects to host servers through a Photonic Fabric NIC (PF-NIC), a PCIe Gen6/CXL 3.1 adapter card that bridges the host's CXL interface to the photonic fabric interconnect. The PF-NIC contains a variant of the PF Memory Module that omits external DDR5 and HBM memory, retaining only the PF ASIC and optical connectivity. While the PF Memory Appliance also supports attachment via an AI accelerator co-packaged with a Photonic Fabric Chiplet (PF Chiplet), this paper focuses on the PF-NIC implementation. Fig. 4 illustrates the connectivity diagram. An External Laser Source (ELS) is required to produce optical illumination for the PF Memory Modules and the PF-NIC. The PF-NIC terminates the CXL link at the host interface, extracts CXL.mem transaction fields from incoming flits, and forwards them as photonic fabric packets to the target PF Memory Module, which services the memory request and returns the response over the same optical link.

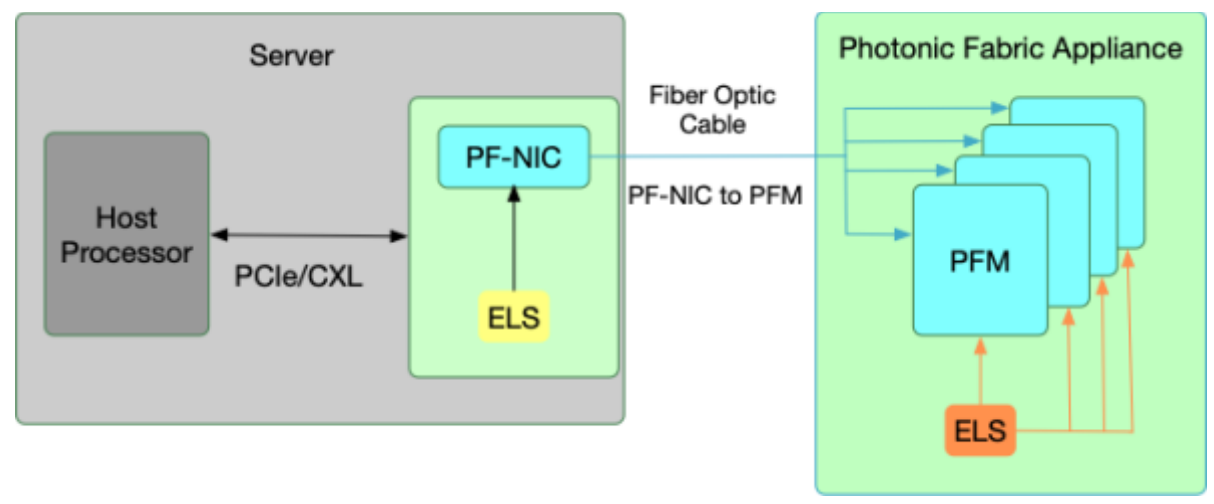

Fig. 4. PF Memory Appliance server connectivity via Photonic Fabric NIC (PF-NIC) over PCIe/CXL interface.

### B. PF Memory Module Design

The advantage of the PF Memory Module compared to other photonic interconnect solutions lies in its thermal stability, bandwidth density, and energy efficiency, achieved through careful selection of photonic components and co-packaging strategies [21].

The PF Memory Module employs Germanium-Silicon (GeSi) Electro-Absorption Modulators (EAMs) rather than Micro-Ring Resonators (MRRs) or Mach-Zehnder Modulators (MZMs). MRRs offer high modulation efficiency but require active wavelength tuning due to thermal sensitivity, adding power overhead [43]. MZMs are thermally stable but require millimeter-scale device lengths. EAMs provide both thermal stability and sub-100 µm footprints [44], enabling tight integration with the ASIC without optical link degradation from thermal gradients.

The PF Memory Module benefits from separated PIC and ASIC fabrication, enabling advanced 4nm/5nm CMOS nodes for electronics while optimizing the PIC for optical performance. This co-packaging approach using standard 2.5D/3D assembly eliminates DSP typically required for long-reach SerDes, allowing the PF Memory Module to achieve significantly lower power consumption than conventional pluggable optics or electrical CXL solutions.

### C. *Advantages over CXL+Switches*

Electrical CXL architectures face three fundamental limitations when scaling to TB-capacity multi-host memory pools. First, multi-host sharing requires CXL switches that introduce 70–500 ns latency per hop plus variable arbitration delays under contention—latency that compounds in the hierarchical topologies necessary for large deployments [16, 30]. Second, signal integrity constraints limit cable reach to several meters, requiring power-hungry retimers for modest distance extension while precluding true rack-scale disaggregation [14]. Third, electrical switch power scales super-linearly with port count and distance, creating thermal density challenges that constrain deployment flexibility..

The PF Memory Appliance's passive fiber shuffle addresses all three limitations through our architectural departure: replacing active electrical switches with dedicated optical paths. The 256-path full-mesh topology eliminates switch arbitration entirely, providing deterministic latency with no contention in the interconnect fabric. Photonic transmission operates over datacenter fiber distances with distance-independent power consumption, enabling rack-scale reach while reducing total interconnect power compared to equivalent electrical deployments. The all-to-all connectivity ensures bandwidth scales linearly with hosts rather than degrading under contention—each host maintains full 128GB/s access regardless of concurrent activity from other hosts.

## IV. Emulation Methodology and Results

### A. *Emulation Framework*

The experimental framework is implemented on the Siemens Veloce Strato-M emulation platform, which employs a processor-based architecture powered by the Crystal accelerator chip. This platform provides deterministic execution and full signal visibility, enabling cycle-accurate debugging and detailed hardware analysis.

All experiments were conducted on a full Strato-M chassis configured with 4 capacity modules and 64 Advanced Board Variants (ABVs). These modules provide sufficient gate capacity to emulate two system-on-chip (SoC) instances: PF-NIC and PF Memory Module, interconnected through 16 photonic fabric (PF) links.

All our design databases for PF-NIC incorporate a CXL memory controller connected to the Siemens VirtuaLab transactor, which emulates a PCIe/CXL Root Complex (RC). In a production system, the RC serves as the central control point of the PCIe/CXL hierarchy, linking the processor and system memory to the PCIe fabric. To emulate the host environment, we use QEMU to model a CPU running a guest Linux operating system that functions as the PCIe/CXL host. C/C++ benchmark programs execute within this environment and generate memory read and write requests to the CXL-attached memory.

Although the CXL memory can be exposed to the host as a NUMA node, this configuration introduces additional operating system overhead, including page table management and other kernel-level memory operations. These activities generate auxiliary transactions that obscure the memory access behavior of the benchmark. To isolate application-generated memory accesses, the benchmark directly maps the CXL memory region using the mmap system call and performs reads and writes through the resulting virtual address range. This approach eliminates extraneous OS-induced traffic and produces significantly cleaner transaction traces in the VirtuaLab protocol analyzer.

### B. *Bandwidth Characterization*

A critical design requirement for the PF Memory Module is that its memory subsystem must maintain full bandwidth and not be a bottleneck to the photonic link, regardless of transfer size, access pattern, or back-pressure conditions. Each Photonic Fabric Memory Module incorporates both HBM stacks, which is configured as a direct-mapped cache, and DDR5 DIMMs serving as main memory. The HBM stack is intentionally provisioned with significantly higher bandwidth than the maximum capacity of the photonic link, ensuring the link always operates at full utilization even under access patterns where DDR5 bandwidth alone would be insufficient to sustain peak throughput.

A single CXL host is limited to 128 GB/s due to the PCIe Gen6 interface constraint and therefore cannot independently saturate a PF Memory Module. To accurately quantify PF Memory Module bandwidth, we measure throughput directly at the PF Memory Module-to-memory channel interface by driving the PF link to saturation. Throughput is captured over a defined observation window to reflect steady-state maximum bandwidth. We execute multiple memory access traces on the emulator across a range of configurations. The traces encompass both linear and random-access address patterns with transfer sizes ranging from 32B to 1KB, designed to stress-test the memory subsystem under diverse access scenarios.

The achievable bandwidth depends critically on the mapping of the PF memory addresses to physical memory addresses. Configurable PF address bits in the design determine which DDR channel, HBM pseudo-channel and channel, and HBM stack are accessed for each transaction, directly controlling how memory traffic is interleaved across memory channels. We identify a fixed set of interleave configuration parameters that distributes traffic uniformly across all memory channels, enabling full bandwidth utilization across the evaluated transfer sizes and access patterns for all 16 PF Memory Module ports.

With the optimized configuration parameters applied, the PF Memory Module achieves 100% bandwidth utilization across all 16 ports, sustaining 25.6 GB/s per port after accounting for protocol and metadata overhead. This throughput is maintained across the full range of transfer sizes (32B–1KB) and access patterns (linear and random) evaluated, confirming that the HBM cache provides sufficient bandwidth headroom to sustain full link utilization under all tested conditions. These bandwidth results were measured on a single PF Memory Module handling traffic representative of 16 concurrent hosts. In a full PF Memory Appliance deployment, CXL host-initiated burst transactions are interleaved across all 16 PF Memory Modules, scaling the aggregate system bandwidth accordingly.

### C. *Latency Characterization*

The end-to-end memory access latency is decomposed into three components: (1) the time required for a QEMU-generated read or write request to reach the Root Complex (RC), (2) the latency within the CXL protocol stack, and (3) the time required for the request to propagate from the CXL protocol interface through the memory subsystem and return with a response. However, the first component is not directly measurable, as delays introduced within QEMU between request generation and arrival at the RC are not observable in the trace.

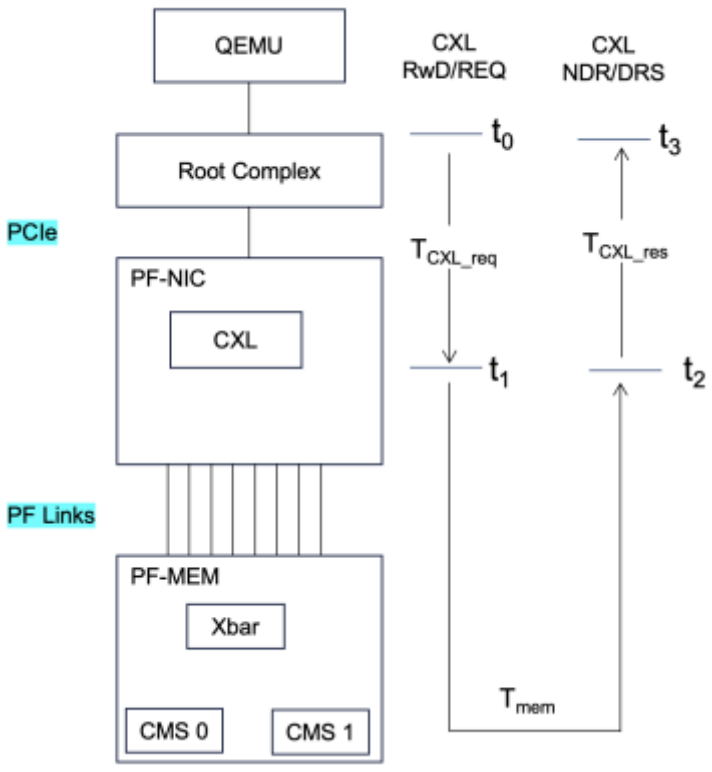

Fig. 5. Block diagram of the latency breakdown.

To quantify the remaining components, we rely on VirtuaLab protocol analyzer traces, which record timestamps at the RC. These timestamps are correlated with waveform signals captured at the CXL protocol interface and at the point where the request exits PF-NIC to extract precise timing points along the request path. Using these measurements, the round-trip latency is expressed as

$$T = T_{CXL_req} + T_{mem} + T_{CXL_resp} \quad (4)$$

where $T_{CXL_req}$ represents the request processing time within the CXL controller, $T_{mem}$ denotes the propagation time from the CXL protocol interface to the memory subsystem and back, and $T_{CXL_resp}$ corresponds to the response processing time in the CXL controller. The intrinsic latency of the CXL controller is estimated using simulation, as simulation provides more accurate cycle-level timing than emulation. Fig. 5 shows the breakdown of the latency captured.

At the protocol level, each CXL write operation is observed as a CXL RwD (Read with Data) transaction followed by a CXL NDR (No Data Response) acknowledgement. Similarly, each CXL read operation appears as a CXL REQ (Request) followed by a CXL DRS (Data Response).

We evaluate latency under two workload conditions. The first configuration, referred to as idle latency, issues a single write transaction followed by a read, capturing system performance under minimal load. The second configuration, referred to as loaded latency, generates a sequence of read and write operations, offset by 64 bytes to remove any cache considerations, that exceeds the implemented CXL credit exchange limits, thereby stressing the protocol pipeline and exposing contention effects.

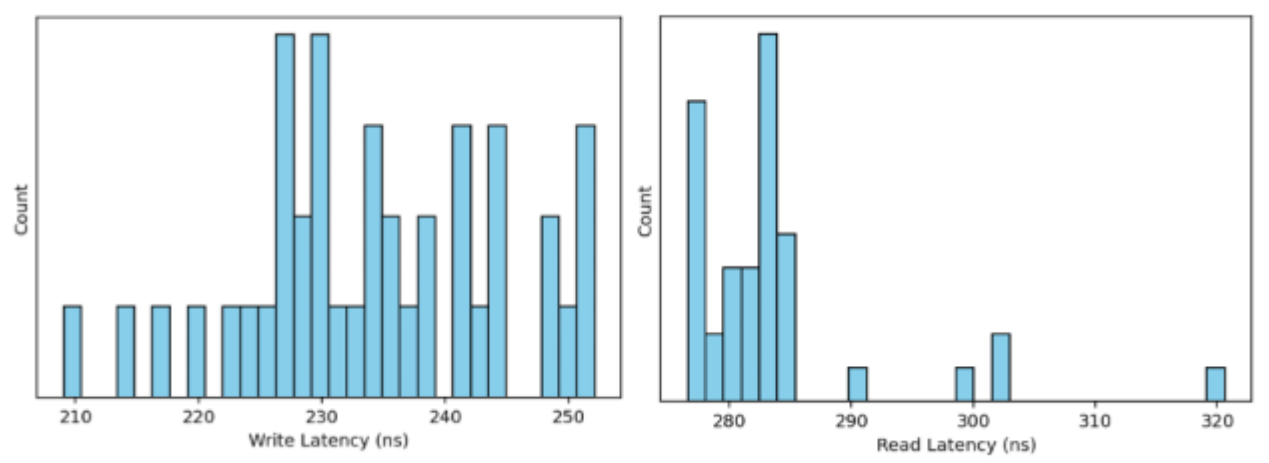

Fig. 6. Latency distribution under loaded conditions for (a) Write and (b) Read operations.

Under the idle latency configuration, we measure the latency components defined in Eq. (2). The memory access component, $T_{mem}$, is measured at 153ns for write operations and 212 ns for read operations, reflecting the expected asymmetry due to data return overhead in reads. To ensure stability, we repeat the experiment multiple times and observe negligible variation with measured values remaining identical or varying at most by 1-2 ns, indicating a highly deterministic execution environment under idle conditions.

From our simulation-based analysis, the CXL controller latencies are estimated for $T_{CXL_req}$ to be 24 ns and $T_{CXL_resp}$ as 47 ns on average. Combining these components, the total round trip latency under idle conditions is:

- Write Latency: 24 + 153 + 47 = 224 ns
- Read Latency: 24 + 212 + 47 = 283 ns

Under the loaded latency configuration, measured latency variability increases as the number of outstanding requests exceeds the CXL credit exchange limits implemented in PF-NIC, resulting in queuing and backpressure within the protocol pipeline.

From Fig. 6, the write latencies are concentrated in the 210 to 250 ns range, with most samples clustered around the mid-range, indicating moderate sensitivity to load. In contrast, read latencies exhibit a wider spread, with majority of samples concentrated around 275 to 290 ns and a pronounced tail extending to 320 ns. This behavior reflects increased contention in read completion path, where data return dependencies increase queuing delays under saturation, leading to tail latency amplification.

For comparison, Yang et al. [16] report an average CPU-to-pool latency of 800–900 ns for 64-byte accesses in their CXL 2.0 switched memory pool (Beluga), measured via Intel MLC. This latency comprises CXL protocol processing at the host, switch forwarding delay, and DDR5 device access. In contrast, our characterization of the CXL pod using PF Memory Appliance yields an average 64-byte access from memory pool to host results in a latency of 350 ns, representing a more than 50% reduction.

## V. Software Integration

This section describes how our optically connected CXL memory appliance is discovered and enumerated with possible integration paths in the ML inference software stack, from OS-level device bring-up through framework-specific KV cache management.

**Device Discovery and OS enumeration**: PF Memory Appliance is a CXL Type-3 (CXL.mem) device. At boot, each host's platform firmware discovers the appliance via PCIe link training and identifies it as CXL-capable through the Designated Vendor-Specific Extended Capability (DVSEC) in the PCIe Configuration Space [22]. The firmware reads the device's Host-managed Device Memory (HDM) capability registers to determine total capacity and interleave granularity, then populates ACPI tables. The CXL Early Discovery Table (CEDT) contains a CXL Host Bridge Structure (CHBS) and a CXL Fixed Memory Window Structure (CFMWS) describing the host-physical address (HPA) range assigned to the appliance [16]. In the multi-host configuration, we ensure that each PFNIC presents a consistent view of the shared memory region within the PF Memory Appliance. Each host independently maps this region into its own physical address space via CFMWS. The Linux CXL subsystem (cxl_acpi, cxl_pci, cxl_mem) validates the CEDT, configures HDM decoders, and surfaces the region as a Device-DAX device (/dev/daxN.M). We retain DAX mode for explicit placement control rather than using it as system-RAM mode.

**Multi-host shared memory.** Each host's inference process maps the DAX device via mmap() and registers the region with CUDA using cudaHostRegister(), pinning pages for GPU DMA access. Since each host may map the region at a different virtual address, all shared data structures in the appliance - the KV block index, allocator metadata, and coordination state - use offset-based addressing relative to the region base rather than virtual pointers.

The appliance supports coherence across hosts using a rendezvous-based consistency protocol. When a producer (prefill worker) writes KV blocks and associated metadata to the shared region, it issues a rendezvous signal through a designated coordination structure in the appliance's memory. Consumer nodes (decode workers) observe the rendezvous before accessing the written data, ensuring that all prior writes by the producer are visible. This approach avoids the overhead of per-cache line flush/fence operations on every metadata write while providing correct cross-node ordering guarantees. Bulk KV data written by GPU DMA does not require CPU-side coherence management, as the DMA completion event on the producer side inherently ensures that data has reached the appliance before the rendezvous signal is published.

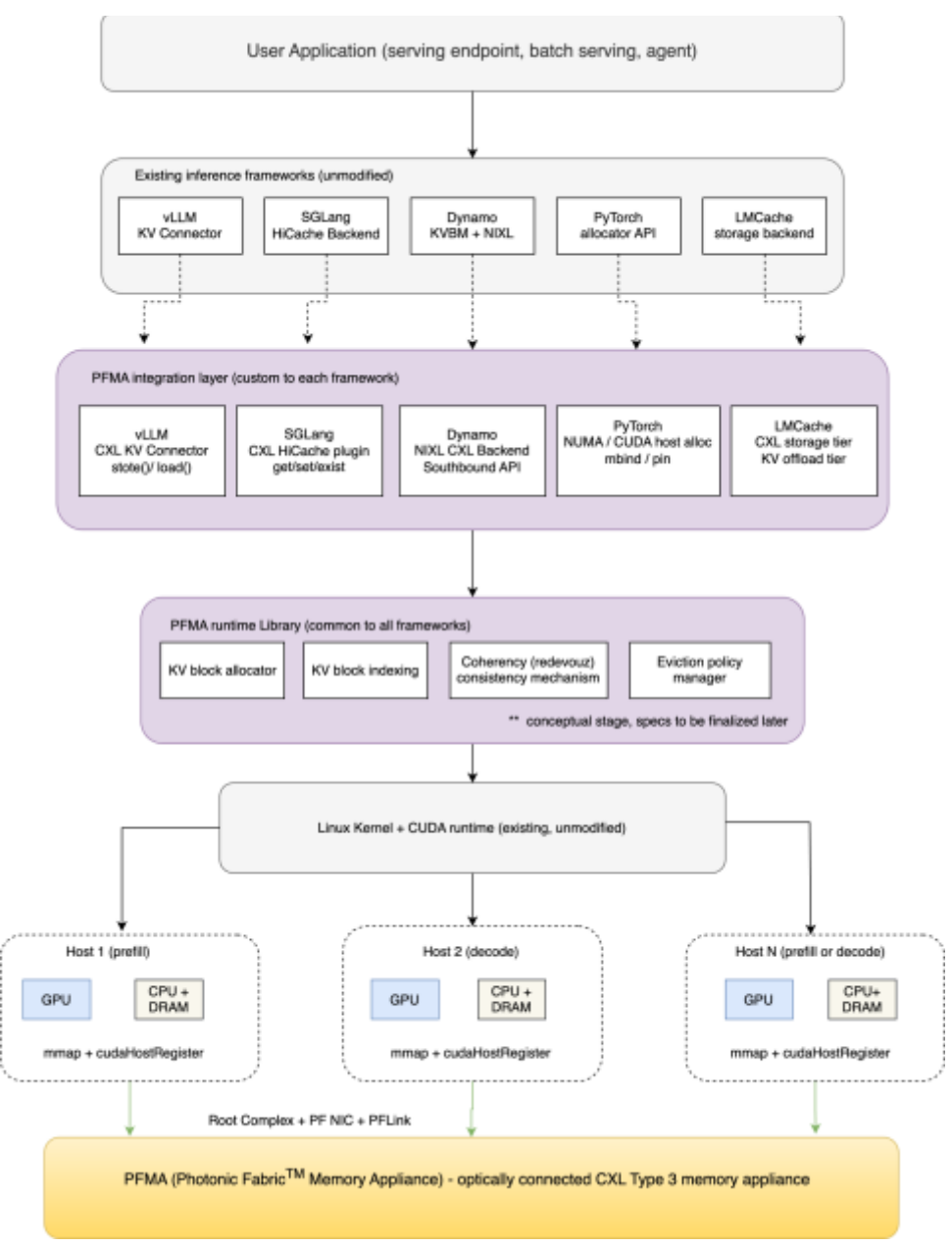

Fig. 7. Software integration approach of PF Memory Appliance with various ML Frameworks

**Inference framework integration.** The PF Memory Appliance can integrate with vLLM, SGLang, and NVIDIA Dynamo through their existing pluggable interfaces, without modifications to engine internals. For vLLM, we will implement a CXL KV connector whose store() and load() methods issue GPU-to-PF Memory Appliance and PF Memory Appliance-to-GPU DMA via cudaMemcpyAsync, with KV block lookup against the shared index [23,24]. For SGLang, a HiCache storage backend can implement the get/exist/set interface over the appliance region, registering it as a tier in HiCache's hierarchical radix tree between GPU HBM and host DRAM [8]. For Dynamo, a NIXL backend plugin can implement the South-Bound API (registerMem, connect, prepXfer, postXfer, checkXfer etc.) [13], registering the appliance as a new memory section type; NIXL's automatic backend selection routes GPU to PF Memory Appliance transfers through this plugin, propagating support to all NIXL-integrated engines (vLLM, SGLang, TensorRT-LLM,LMCache) without per-framework modifications. At the PyTorch level, the appliance is accessible either as a NUMA-aware CPU allocator via mbind (MPOL_BIND) for per-tensor placement of latency-tolerant data [25], or as a CUDAHost allocator for direct GPU DMA. Pytorch currently does not support heterogenous memory tiers. Memory in Pytoch are limited to: device memory (on chip DRAM) and system host memory. An active PyTorch RFC proposes native memory-type annotations in torch.device that would further simplify this integration [26] where PF Memory Appliance can be added as a new memory type for computational devices natively supported by PyTorch.

Fig. 7 shows the software integration architecture and how the PF Memory Appliance integrates with existing ML frameworks. Gray components are existing and unmodified; purple components are new. Framework-specific connectors (top purple tier) implement thin adapters against each engine's pluggable API. The common runtime library (bottom purple tier) encapsulates shared-memory management, rendezvous-based consistency, and offset-addressed KV indexing across all hosts. Multiple hosts access the appliance over optical CXL with GPU DMA for KV payloads and CPU load/store for metadata.

## VI. Performance Evaluation

### A. Methodology

We combine the empirical characterization from Section II-B with the emulation results from Section IV to construct a performance model that quantifies the end-to-end benefits of PF Memory Appliance for production LLM serving workloads.

#### 1) Hardware Configurations

To evaluate the impact of PF Memory Appliance on LLM serving, we compare three hardware configurations, all configurations serve a LLaMA 405B model:

Baseline: 8×H200 DGX system with 2 TB of host DRAM.

PF Memory Appliance-augmented: 8×H200 DGX system with 2 TB of host DRAM and a CXL-connected PF Memory Appliance providing 32 TB of shared pooled memory at 128 GB/s unidirectional bandwidth per host.

#### 2) Workload Description

We design a multi-turn conversation workload to simulate chatbot-style production deployments such as customer support and interactive assistants. This workload faces a common challenge on current hardware: it either requires additional server clusters to meet latency SLOs or suffers degraded TTFT as concurrent request volume increases. Each conversation consists of 10 turns, with each turn comprising 5,120 input tokens and 500 output tokens. A 512-token system prompt prefix is shared across all conversations, creating prefix-sharing opportunities that exercise KV cache reuse. We vary the total number of conversations to control workload intensity, reflecting increasing concurrent user load in production. We measure mean TTFT, cache hit ratio across different conversation counts for each hardware configuration.

#### 3) Simulation Framework

We evaluate our proposed architecture using LLMServingSim [27], a runtime-driven HW/SW co-simulation framework for LLM inference serving that captures the dynamic, autoregressive nature of LLM serving by simulating each decoding iteration individually through a loop of batch scheduling, hardware execution, and system pipeline modeling. The framework supports continuous batching with dynamic batch composition, prefill-decode disaggregation, and—critically for our study—a configurable multi-tier memory hierarchy that models device memory, host DRAM, CXL-attached memory as distinct tiers with independently configurable bandwidth and latency parameters. The simulator employs profile-based operator modeling: a one-time profiling pass on target hardware (8×H200 for LLaMA 405B model in our case) captures per-layer compute latencies, which are then replayed within the runtime-driven simulation loop, achieving 1.9% average error against real serving deployments.

We leverage these capabilities to model our baseline 8×H200 DGX system and to introduce PF Memory Appliance as an additional memory tier connected via CXL, configuring its bandwidth and capacity parameters to match the experimentally characterized photonic fabric specifications from Section IV.

### B. Results

Fig. 8(a) shows mean TTFT as the total token working set scales from 50 conversations to 300 conversations. With the 32 TB PF Memory Appliance, TTFT remains flat at 2,690 ms regardless of conversation count—all KV cache entries are retained in the pooled memory, and subsequent turns benefit from full prefix cache hits (82% hit rate). In contrast, the 2 TB baseline degrades rapidly: at 100 conversations, mean TTFT rises to 8,156 ms (3× higher) as the LRU eviction policy forces full re-computation of evicted KV cache. At 300 conversations, the gap widens to 6.6×, with the prefix hit rate collapsing from 82% to 6.35% (Fig. 8(b)).

The key benefit of CXL-attached PF Memory Appliance is the combination of sufficient capacity and bandwidth in a single memory tier. With 32 TB of pooled memory, a single 8×H200 system can serve 64× more concurrent multi-turn conversations or long-context sessions compared to a conventional 2 TB host DRAM configuration, with zero cache eviction and consistently low TTFT at scale. These results validate the capacity-bandwidth gap identified in Section II-B and demonstrate that PF Memory Appliance effectively eliminates it for production-scale LLM serving workloads.

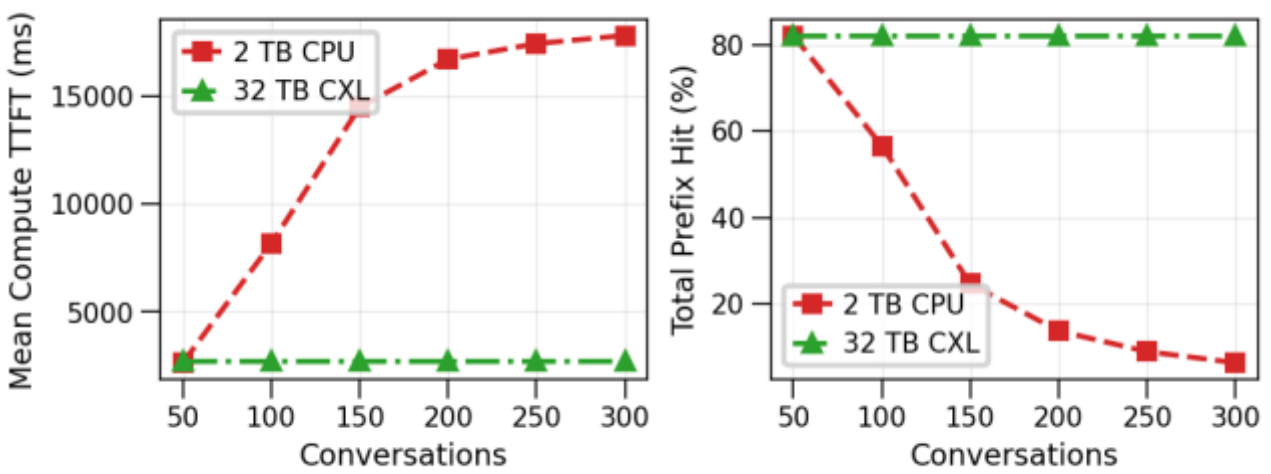

Fig. 8. (a) Mean TTFT results for multi-turn conversation workload over different number of conversations with and without PF Memory Appliance; (b) Total prefix cache hit rate over different number of conversations with and without PF Memory Appliance.

## VII. Related Work

### A. KV Cache Management Systems

PagedAttention [6] established block-based KV management inspired by OS virtual memory, reducing fragmentation from 60–80% to under 4%. SGLang's RadixAttention [28] introduced automatic prefix reuse via radix trees, and its HiCache extension [8] added hierarchical tiering across GPU, CPU DRAM, and distributed storage with up to 6× throughput and 80% TTFT reduction. NVIDIA Dynamo's KVBM [5] implements the most complete tiered hierarchy with KV-aware routing and NIXL [13] for cross-node transfers. LMCache [7] provides middleware abstracting heterogeneous backends with content-hash-based deduplication. Mooncake [11] treats KV cache as a first-class disaggregated resource across cluster-wide CPU DRAM and SSDs, but remains bounded by RDMA bandwidth (~15 GB/s per NIC).

Complementary techniques address cache efficiency from other angles. CacheBlend [29] selectively recomputes 5–18% of tokens per layer to enable non-prefix reuse for RAG workloads.

CacheGen [30] compresses KV cache into 3.5–4.3× smaller bitstreams for network streaming. InfiniGen [31] speculatively prefetches critical KV entries from CPU memory. KVQuant [32] provides sub-4-bit quantization enabling 10M+ token contexts. FlashAttention [33, 34] optimizes attention computation by minimizing GPU SRAM↔HBM data movement. These optimizations are orthogonal to memory pooling and compose directly with PF Memory Appliance.

The common limitation is that per-node memory capacity forces multi-tier caching with escalating latency penalties, and cross-node sharing requires network transfers with unpredictable overhead—the gap the PF Memory Appliance addresses.

### B. *CXL Memory Disaggregation*

Pond [30] showed that 8–16 socket CXL pools capture most DRAM savings but estimated 70–90 ns overhead rising to >180 ns at rack scale. DirectCXL [35] demonstrated 8.3× lower latency than RDMA via load/store semantics. TPP [36] contributed transparent page placement for tiered CXL memory, now merged into the Linux kernel. Melody [37] found significant latency instability across commercial CXL devices, with p99.99 tail latency exceeding 700 ns–1 µs. COAXIAL [38] proposed replacing all DDR with CXL for 1.39× average speedup. Seo et al. [25] explored CXL-attached memory for long-context LLM fine-tuning.

The fundamental electrical CXL constraints are well-documented: copper cables limited to ≤2 m, 50–70 ns per switch hop, and power-hungry retimers for distance extension. Hybrid approaches like Rcmp [39] (CXL intra-rack + RDMA inter-rack) and DFabric [40] work around but do not solve these constraints. The PF Memory Appliance's passive fiber shuffle directly addresses all three: extending reach to 100+ m, eliminating switch-hop latency, and reducing interconnect power.

### C. *Photonic Interconnects in Datacenter*

Co-packaged optics has reached production maturity, driven by the need to eliminate DSP latency (~100–150 ns) and power (~15–20 pJ/bit) from pluggable transceivers. Broadcom's Tomahawk 6 ships with integrated optical engines at ~5 pJ/bit [41]. Intel's OCI achieves 4 Tbps at 5 pJ/bit with 100 m reach [43], explicitly targeting memory disaggregation.

GeSi electro-absorption modulators are emerging as the preferred CPO technology. Recent demonstrations include >100 GHz bandwidth at sub-100 µm active lengths [43] and 400 Gb/s per lane on 300 mm silicon photonics [44]. EAMs avoid MRR thermal stabilization overhead and footprint penalties—the approach used in the PF Memory Module [17, 21].

Lightmatter's Passage M1000 [45] integrates 34 chiplets with 114 Tbps optical bandwidth. Ayar Labs' TeraPHY [46] delivers 8 Tbps as the first UCIe optical chiplet supporting CXL protocol transport. NVIDIA's SN6800 Spectrum-X switch uses an integrated fiber shuffle between multiple ASICs [47], validating the passive optical shuffle concept in production. The PF Memory Appliance builds on these advances by combining CXL load/store semantics with a passive 16×16 fiber shuffle for switch-free, rack-scale memory pooling.

## VIII. Limitations and Future Work

This study has several limitations. First, our empirical characterization is focused on disk storage experiments on the 8B and 405B models; more characterization of the MoE variants, including new architectures from DeepSeek and Qwen models, will be future work. Second, the PF Memory Appliance performance projections in the serving evaluation are derived from emulation-characterized bandwidth and latency parameters fed into the LLMServingSim framework; validation on the physical PF Memory Appliance hardware with end-to-end inference workloads remains pending.

Future work will address these gaps through several directions. We plan to conduct end-to-end validation on the PF Memory Appliance hardware once it becomes available later this year, including the integration with the vLLM and SGLang connectors described in Section V. We will extend the serving evaluation to multi-GPU, multi-host configurations to quantify the benefits of the PF Memory Appliance's shared memory pool for cross-host KV reuse in disaggregated inference deployments. Cost-benefit analysis comparing PF Memory Appliance against alternative scaling approaches—including CXL switch hierarchies, active optical interconnects, and simply provisioning additional servers—will provide additional deployment guidance. Finally, we plan to explore PF Memory Appliance's applicability beyond KV cache management, including model weight sharing across hosts and checkpoint storage for fault-tolerant serving.

## IX. Conclusion

We presented a comprehensive characterization of KV cache retrieval across the memory hierarchy to reveal a fundamental capacity-bandwidth gap in current systems. Host memory achieves up to 100× speedup over re-computation but supports only tens of concurrent long-context users, while SSD storage scales in capacity but delivers at most 9.7× speedup. Electrical CXL memory pooling can theoretically bridge this gap but faces switch latency, reach, and power constraints that prevent practical TB-scale deployments.

The Photonic Fabric Memory Appliance addresses these limitations through a passive optical fiber shuffle that replaces electrical CXL switches, delivering 32 TB of shared DDR5 memory across 16 hosts at 128 GB/s per host with a switch-free full-mesh topology. Emulation results demonstrate over 50% latency reduction compared to electrical CXL switched pools, with full-link bandwidth utilization across diverse access patterns. Serving simulations demonstrate that PF Memory Appliance eliminates capacity-induced eviction cliffs, sustaining flat TTFT as workload scales while baseline configurations degrade rapidly once their memory capacity is exceeded. At 300 concurrent multi-turn conversations, the PF Memory Appliance achieves 6.6× lower TTFT than the baseline when the baseline's memory is fully exhausted. These results demonstrate that the photonic-CXL memory pooling effectively closes the capacity-bandwidth gap for production-scale LLM serving.

## References

[1] Anthropic, "Claude model family," 2025. [Online]. Available: https://www.anthropic.com/claude

[2] Google DeepMind, " Gemma 3 Technical Report," 2025. [Online]. Available: https://arxiv.org/abs/2503.19786

[3] OpenAI, "GPT-5.4," 2025. [Online]. Available: https://chatgpt.com/overview?openaicom_referred=true

[4] Meta, "LLaMA 4 model family," 2025. [Online]. Available: https://ai.meta.com/blog/llama-4-multimodal-intelligence/

[5] NVIDIA, "How to reduce KV cache bottlenecks with NVIDIA Dynamo," NVIDIA Technical Blog, 2025. [Online]. Available: https://developer.nvidia.com/blog/how-to-reduce-kv-cache-bottlenecks-with-nvidia-dynamo/

[6] W. Kwon, Z. Li, S. Zhuang, Y. Sheng, L. Zheng, C. H. Yu, J. E. Gonzalez, H. Zhang, and I. Stoica, "Efficient memory management for large language model serving with PagedAttention," in Proc. ACM SOSP, 2023. https://doi.org/10.1145/3600006.3613165

[7] Y. Liu, H. Li, Y. Du, S. Song, J. Zhu, J. Peng, S. Chen, K. Xiong, and J. Lu, "LMCache: An efficient KV cache layer for enterprise-scale LLM inference," arXiv:2510.09665, 2025. https://doi.org/10.48550/arXiv.2510.09665

[8] LMSYS, "SGLang HiCache: Fast hierarchical KV caching with storage backends," 2025. [Online]. Available: https://lmsys.org/blog/2025-09-10-sglang-hicache/

[9] Y. Zhong, S. Liu, J. Chen, J. Hu, Y. Zhu, X. Liu, X. Jin, and H. Zhang, "DistServe: Disaggregating prefill and decoding for goodput-optimized large language model serving," in Proc. USENIX OSDI, 2024. https://doi.org/10.48550/arXiv.2401.09670

[10] P. Patel, E. Choukse, C. Zhang, A. Shah, Í. Goiri, S. Maleki, and R. Bianchini, "Splitwise: Efficient generative LLM inference using phase splitting," in Proc. IEEE/ACM ISCA, 2024. https://doi.org/10.1109/ISCA59077.2024.00019

[11] R. Qin et al., "Mooncake: Trading more storage for less computation—a KVCache-centric architecture for serving LLM chatbot," ACM Trans. Storage, 2025. https://doi.org/10.1145/3773772

[12] NVIDIA, "GPUDirect RDMA documentation." [Online]. Available: https://docs.nvidia.com/cuda/gpudirect-rdma/

[13] NVIDIA, "NIXL: NVIDIA Inference Transfer Library," 2025. [Online]. Available: https://github.com/ai-dynamo/nixl

[14] CXL Consortium, "Compute Express Link Specification 3.1," 2024. [Online]. Available: https://www.computeexpresslink.org/download-the-specification

[15] J. Yoon et al., "TraCT: Disaggregated LLM serving with CXL shared memory KV cache at rack-scale," arXiv:2512.18194, 2025. https://doi.org/10.48550/arXiv.2512.18194

[16] Z. Yang et al., "Beluga: A CXL-based memory architecture for scalable and efficient LLM KVCache management," arXiv:2511.20172, 2025. https://doi.org/10.48550/arXiv.2511.20172

[17] G. Balamurugan et al., "A 56-Gb/s hybrid silicon photonic and 5-nm CMOS 3-D-integrated transceiver for optical compute I/O," IEEE J. Solid-State Circuits, 2026.

[18] S. Sahni et al., "Photonic Fabric for memory and compute disaggregation," in Proc. OFC, paper W3D.1, 2025. https://doi.org/10.1364/OFC.2025.W3D.1

[19] S. Potluri, K. Hamidouche, D. Bureddy, and D. K. Panda, "Benchmarking GPUDirect RDMA on modern server platforms," NVIDIA Technical Blog, 2014. [Online]. Available: https://developer.nvidia.com/blog/benchmarking-gpudirect-rdma-on-modern-server-platforms/

[20] llm-d Project, "Distributed KV cache scheduling and offloading," 2025. [Online]. Available: https://llm-d.ai

[21] P. Winterbottom et al., "Photonic Fabric: Co-packaged photonic interconnect technology," Celestial AI, 2025.

[22] Intel, "CXL Type 3 Memory Device Software Guide," Document 643805, Rev 1.1. [Online]. Available: https://www.intel.com/content/www/us/en/content-details/643805/

[23] vLLM Project, "KV Connector API." [Online]. Available: https://docs.vllm.ai

[24] vLLM Blog, "Inside vLLM's new KV offloading connector," Jan. 2026. [Online]. Available: https://blog.vllm.ai

[25] S. Seo et al., "CXL-attached memory allocation for long-context LLM fine-tuning," arXiv:2507.03305, 2025. https://doi.org/10.48550/arXiv.2507.03305

[26] PyTorch, "Intra-device heterogeneous memory allocation support," RFC #153745, 2025. [Online]. Available: https://github.com/pytorch/pytorch/issues/153745

[27] S. Cho, H. Choi, G.-I. Kwon, L. Kim, and M. Rhu, "LLMServingSim: A runtime-driven HW/SW co-simulation framework for LLM inference serving," in Proc. IEEE IISWC, 2024; extended in IEEE Comput. Archit. Lett., 2025. https://doi.org/10.1109/IISWC63097.2024.10763697

[28] L. Zheng et al., "SGLang: Efficient execution of structured language model programs," arXiv:2312.07104, 2024. https://doi.org/10.48550/arXiv.2312.07104

[29] J. Yao, H. Li, Y. Liu, S. Song, K. Zhang, Y. Xiong, and J. Lu, "CacheBlend: Fast large language model serving for RAG with cached knowledge fusion," in Proc. ACM EuroSys, 2025. https://doi.org/10.1145/3689031.3696098

[30] H. Li et al., "Pond: CXL-based memory pooling systems for cloud platforms," in Proc. ACM ASPLOS, 2023. https://doi.org/10.1145/3575693.3578835

[31] J. Lee et al., "InfiniGen: Efficient generative inference of large language models with dynamic KV cache management," in Proc. USENIX OSDI, 2024. [Online]. Available: https://www.usenix.org/conference/osdi24/presentation/lee

[32] C. Hooper, S. Kim, H. Mohammadzadeh, M. W. Mahoney, Y. S. Shao, K. Keutzer, and A. Gholami, "KVQuant: Towards 10 million context length LLM inference with KV cache quantization," in Proc. NeurIPS, 2024. https://doi.org/10.48550/arXiv.2401.18079

[33] T. Dao, D. Fu, S. Ermon, A. Rudra, and C. Ré, "FlashAttention: Fast and memory-efficient exact attention with IO-awareness," in Proc. NeurIPS, 2022. https://doi.org/10.48550/arXiv.2205.14135

[34] T. Dao, "FlashAttention-2: Faster attention with better parallelism and work partitioning," in Proc. ICLR, 2024. https://doi.org/10.48550/arXiv.2307.08691

[35] D. Gouk, S. Lee, M. Kwon, and M. Jung, "DirectCXL: Direct memory access for CXL-enabled memory disaggregation," in Proc. USENIX ATC, 2022. [Online]. Available: https://www.usenix.org/conference/atc22/presentation/gouk

[36] H. Maruf, H. Wang, A. Dhakal, H. Li, Z. Cai, M. Mosharaf, and K. Akeley, "TPP: Transparent page placement for CXL-enabled tiered memory," in Proc. ACM ASPLOS, 2023. https://doi.org/10.1145/3582016.3582063

[37] J. Liu et al., "Melody: Systematic CXL memory characterization and performance analysis at scale," in Proc. ACM ASPLOS, 2025. https://doi.org/10.1145/3676641.3715987

[38] A. Cho et al., "COAXIAL: A CXL-centric memory system for scalable servers," in Proc. IEEE/ACM SC, 2024. https://doi.org/10.1109/SC41406.2024.00101

[39] Z. Wang et al., "RCMP: Reconstructing RDMA-based memory disaggregation via CXL," ACM Trans. Archit. Code Optim., 2024. https://doi.org/10.1145/3634916

[40] DFabric, "Scaling out data parallel applications with CXL-Ethernet hybrid interconnects," arXiv:2409.05404, 2024. https://doi.org/10.48550/arXiv.2409.05404

[41] Broadcom, "Tomahawk 6 (Davisson) co-packaged optics switch," 2025. [Online]. Available: https://www.broadcom.com/products/ethernet-connectivity/switching/strataxgs/bcm78900

[42] Intel, "Optical Compute Interconnect (OCI)," demonstrated at OFC, 2024. [Online]. Available: https://www.intel.com/content/www/us/en/research/optical-compute-interconnect.html

[43] L. Steckler et al., "Monolithic electro-optic platform on silicon with bandwidth of 100 GHz and beyond," Nature Communications, Nov. 2025. https://doi.org/10.1038/s41467-025-66566-2

[44] imec, "Beyond-110 GHz C-band GeSi electro-absorption modulator on 300 mm silicon photonics," presented at ECOC, 2025.

[45] Lightmatter, "Passage M1000 photonic superchip," 2025. [Online]. Available: https://lightmatter.co/products/passage/

[46] Ayar Labs, "TeraPHY: 3rd generation UCIe optical chiplet," presented at OFC, 2025. [Online]. Available: https://ayarlabs.com/teraphy/

[47] NVIDIA, "A new era in data center networking with NVIDIA silicon photonics-based network switching," NVIDIA Technical Blog, 2025. [Online]. Available: https://developer.nvidia.com/blog/a-new-era-in-data-center-networking-with-nvidia-silicon-photonics-based-network-switching/